\documentclass[twocolumn,showpacs,preprintnumbers,amsmath,amssymb,rmp]{revtex4}

\usepackage{graphicx}
\usepackage{dcolumn}
\usepackage{bm}
\usepackage{natbib}

\begin{document}

\title{\bf Ocular dominance patterns and the wire length minimization: a numerical study}

\author{Alexei A. Koulakov and Dmitri B. Chklovskii}
\address{Cold Spring Harbor Laboratory, Cold Spring Harbor, NY 11724}

\date{April 8, 2002}

\begin{abstract}
We study a mathematical model for ocular dominance patterns (ODPs) in primary visual cortex. 
This model is based on the premise that ODP is an adaptation to minimize the length of intra-cortical wiring. 
Thus we attempt to understand the existing ODPs by solving a wire length minimization problem. 
We divide all the neurons into two classes: left- and right-eye dominated. 
We find that segregation of neurons into monocular regions reduces wire length if 
the number of connections to the neurons of the same class (intraocular) differs from the number of interocular connections. 
The shape of the regions depends on the relative fraction of neurons in the two classes. 
We find that if both classes are almost equally represented, the optimal ODP consists of interdigitating stripes. 
If one class is less numerous than the other, the optimal ODP 
consists of patches of the less abundant class surrounded by the neurons of the other class. 
We predict that the transition from stripes to patches occurs when the fraction of neurons dominated by the underrepresented eye is about $40\%$. 
This prediction agrees with the data in macaque and Cebus monkeys. 
We also study the dependence of the periodicity of ODP on the parameters of our model.
\end{abstract}

\maketitle

\section{INTRODUCTION}

In the primary visual area (V1) of many mammals, most neurons respond
to the stimulation of two eyes unevenly: they are either right 
or left eye dominated. In some species, right/left eye dominated
neurons are segregated and form a system of alternating monocular regions
 known as the ocular dominance pattern (ODP) (Wiesel and Hubel, 1965, 1969).
In others, ODP is not observed (see Horton and Hocking, 1996b for a comprehensive list of species). 
ODPs, when observed, vary significantly between different species and even between 
different parts of V1 in the same animal.

Most modeling studies of ODP (Erwin et al., 1995; Swindale, 1996) have addressed its development.  
They succeeded in generating ODPs of realistic appearance. 
However, several {\em why} rather than {\em how} questions remained unanswered. For instance, 
(1) why, from functional point of view, do the ODPs exist? 
(2) Why do some mammalian species exhibit ODPs while others do not (Horton and Hocking, 1996b; Livingstone, 1996)? 
(3) Why do the monocular regions have different appearance (stripes as opposed to patches) between
different parts of V1 within the same animal (LeVay et al., 1985)?

The question of functional significance of ODPs has been addressed 
theoretically using the wiring economy principle 
(Mitchison, 1991; Chklovskii, 2000).            
The idea is that evolutionary pressure to keep the brain volume to a
minimum requires making the wiring (axons and dendrites) as short as
possible, while maintaining neuronal functional properties 
(Cajal, 1995; Allman and Kaas, 1974; Cowey, 1979;
Cherniak, 1992; Young, 1992; Chklovskii et al., 2001; Koulakov and Chklovskii, 2001). 
In many cases these functional properties are specified by 
the rules of establishing connections between neurons, or wiring rules.
The problem presented by the wiring economy
principle is therefore to find, for given wiring rules, the spatial neuronal layout
that minimizes the total connection length.
This approach allows to understand many features in cortical maps, such as 
orientation preference maps (Koulakov and Chklovskii, 2001), 
as evolutionary adaptations, which minimize the total cortical volume.

The goal of this study is to find the simplest model, which on one hand
is supported by experimental evidence, and on the other encompasses
most of OD phenomenology. The use of the simple model allows us
to explore its parameter space completely and to give answers to the set of questions above. 
We also evaluate the dependence of the ODP period on the parameters of our model and compare it to the ODP periodicity observed in macaque monkey.
We find that the experimentally observed variation of the period is in agreement with the wiring economy theory.


\section {MODEL AND METHODS}

\subsection{Description of the model}      
 
For the purposes of minimizing the cortical wiring we                           
consider only intra-cortical connections since they constitute
the majority of gray matter wiring (LeVay and Gilbert, 1976; Peters and Payne, 1993; Ahmed et al., 1994). 
We therefore disregard the thalamic afferents and other extra-cortical projections. 
In an attempt to make wiring economy argument more quantitative,
we propose a model describing the component of intracortical circuitry sensitive to OD.                             
The principal elements of our model are therefore the connection rules between cortical neurons. 
To assess the sensitivity of the intracortical wiring to OD 
we examine the connections in the cortical layer $4C\beta$, where OD
is most strongly pronounced. Such sensitivity has been studied
by Katz et al. (1989). They made three observations regarding the wiring rules:

{\it i)} Neurons in the layer $4C\beta$ near the interface between two OD columns arborize
more in home rather than in the opposite eye column. Therefore neurons 
establish more connections with the neurons dominated by the same rather than by 
the opposite eye.

{\it ii)} Axons and dendrites of these neurons have a tendency to bend away from the interface 
between OD columns. This implies that not only they avoid penetration to the opposite OD
column but also they attempt to maintain sufficiently high number of connections
in the home column.

{\it iii)} Axons or dendrites penetrating through the opposite eye
column to the next same eye column are {\em never} observed.
This means that retinotopy has little effect on connections in layer $4C\beta$.  
Indeed the neurons on the edges of two same eye columns separated
by one opposite eye column have on average receptive fields centered next 
to each other. If connections in $4C\beta$ were sensitive to the retinotopic
coordinates, these two edges should be connected (Mitchison, 1991). However out
of 21 cells examined Katz et al. (1989) observed {\em none}
producing axons reaching the next same eye domain. The only
possibility for such cells to be connected is due to the overlap
between dendritic and axonic arbors of two cells 
separated by more than $500$ $\mu$m.                                 
Such possibility is small because of the strong repulsion of the
connections by the opposite eye column located between two cells (observation {\it i)}).


These three observations lay the basis of our model which we now describe.
The elementary unit of our model mimics the columnar organization of the cortex (Mountcastle, 1957) 
and uniformity of ODP along the direction normal to the slab. 
The elementary unit is therefore a microcolumn, which is defined
as a box, spanning the cortex perpendicular to its surface,
whose other two dimensions are smaller than the characteristic scale of 
ODP ($\approx 500\mu$), and yet large enough to include many neurons.
A possible choice of dimensions for such a microcolumn is
{\it thickness of cortex } $(\approx 1.5$mm $) \times 30\mu$ $\times 30\mu$, in which
case it includes about 310 cells in V1 (Rockel et al., 1980). 
The microcolumn units are therefore arranged on a square lattice with 30$\mu$ period.          

Although the choice of the elementary unit size may seem
arbitrary, the results of our calculation are independent of the choice.
The size of the unit is analogous to the integration step, 
which does not affect the value of an integral significantly if
chosen to be small enough.

%
%
\begin{figure}
\centerline{
\includegraphics[width=1.5in]{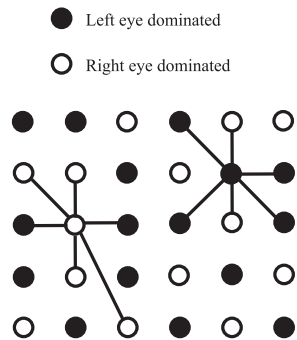}
} 
\caption{   
\protect{Our model illustrated. The units dominated by the left and right eyes are shown by the full and empty circles respectively.
Each unit is {\it required} to make $N_s$ connections to the units of the same OD and $N_o$ connections to the units of the opposite OD.
In this illustration $N_s=4$ and $N_o=2$. The connections satisfying these rules are shown for two units, right and left eye dominated.
Small numbers of connections are chosen for the ease of illustration; in actual implementation of the model both $N_s$ and $N_o$
are large (see below).}
\label{cr}
}
\end{figure}

Motivated by the second observation in layer $4C\beta$ listed above, i.e. that neurons 
maintain a fixed number of connections in the home OD column, we make the following assumption
about the connection rules. Each microcolumn unit must establish 
connections with $N_s$ distinct units dominated by the same eye and $N_o$ units 
dominated by the opposite eye. These connection rules are illustrated in Fig.~\ref{cr}.

Only the relative values of $N_s$ and $N_o$ (rather than absolute) are important 
because of the  arbitrariness in the definition of microcolumn. 
Thus, if a $30\times 30\mu$ microcolumn receives $N_s=10^4$ projections
from the same OD column and $N_o=10^3$ projections from the opposite OD column, 
a $60\times 60\mu$ microcolumn receives four times less projections respectively, or $N_s=2.5 \cdot 10^3$ and $N_o=2.5\cdot 10^2$. 
This is because with coarser units each projection is more effective: connecting to one
$60\times 60\mu$ unit implies connecting to four $30\times 30\mu$ units.
Both these implementations of the model produce the same OD pattern, discretized
in ``pixels'' of different size. The important quantity, which is invariant with respect to the
change of ``pixel''/microcolumn size, is the ratio between $N_s$ and $N_o$ (equal to 10 in this example).
This is the first parameter of our model.

The second parameter of our model is the filling fraction of the units (microcolumns) dominated by the left eye afferents $f_L$ with respect to
the total number of units, averaged over several ODP periods. 
This parameter is the fraction of the left eye dominated units $f_L$ by $f_L+f_R=1$. 
For the majority of important cases $f_L=f_R=1/2$, however on the periphery of visual field one of the
eyes (ipsilateral) is underrepresented. Therefore, its filling fraction is 
less than $1/2$.

The third observation above
implies that the component of connections sensitive to OD is not sensitive to
the retinotopy, and both numbers $N_s$ and $N_o$ do not depend on the
position of the receptive field of the unit. This may be due to significant
scatter of the receptive field of the cells within cortical column 
on the scales of about 1mm (Hubel and Wiesel, 1974).
The position of the receptive field of the microcolumn is therefore
vaguely defined and cannot affect OD sensitive connections significantly.

\subsection{Methods}                               

Given these wiring rules we look for an optimal layout of the microcolumn units
which minimizes the total length of connections.
To find the layout minimizing the total wirelength 
we employ a combination of computational and analytical techniques.
To make our choice of methods clear we first comment on the
expected properties of the solution.

A possible solution of our model is the {\em Salt and Pepper}
layout in which the units dominated by right and left eyes are
uniformly intermixed. In this layout the units belonging
to different eyes are not segregated, ODP is not formed,
and the local values of the filling fraction are equal to $1/2$
(by local value we understand an average over a domain including many
units yet small compared to the period of ODP).
It should be contrasted to the case when units dominated by the same eye fill in large
domains i.e. form the ODP. In the latter case the local values of the filling factor of each
eye vary from 0 to 1. However one can imagine an intermediate
situation when the local filling fraction varies from $1/2-a$ to $1/2+a$,
where the amplitude of variation $0 < a \ll 1/2$. This corresponds to the case 
of {\it weak segregation} into ODP. The weak segregation is
found in squirrel monkey where ODP has fuzzy appearance and until recently was suspected not to be formed 
(Horton and Hocking, 1996).
If $a=1/2$, i.e. the local filling fraction varies from $0$ to $1$,
the ODP's have sharp appearance. Using the general terminology
from binary mixtures (cite diblock copolimer paper) we call this
regime the {\it strong segregation} limit.

The methods useful in the strong segregation limit are not good
in the weak segregation case and vice versa. We use the 
simulated annealing to find the optimum phases for the 
strong and nearly strong segregation cases. Having found the optimum phase
in the strong segregation case to assess the period of ODP  we use the 
exact enumeration technique, which compares layouts belonging to the same phase
with different periods. The treatment of the weak segregation case
requires the use of continuous variables and is done employing the perturbation
theory. Below we describe these methods in more detail.
                            
\subsubsection{Simulated annealing}

The parameters of Metropolis Monte-Carlo method (Metropolis et al., 1953) 
are optimized to render most consistent results for multiple restarts.
We use square $20\times 20$ array of units with periodic boundary
conditions. The units are either left or right eye dominated.
At each step the algorithm attempts to change the dominance of one unit to the
opposite. 
The value of the average filling fraction $f_{L0}$ is enforced by
adding the following term to the total connection length:
\begin{equation}
\delta L = 20.0 L \frac{(f_L-f_{L0})^2}{1/f_{L0}+1/(1-f_{L0})},
\end{equation}
where $L$ and $f_L$ are the current values of the total wirelength and average
filling fraction. Such term in the functional keeps the
current value $f_L$ close to the required value $f_{L0}$.

To map out the phase diagram the values of $f_L$ change from 
$0.2$ to $0.8$ in $0.02$ increments. The values of $N_s$ and $N_o$
satisfy the condition $N_s+N_o=30$ and are changed in unit increments,
i.e. have the following values: 12, 18; 13, 17; 14, 16; 15, 15; 16, 14; etc.
The phases at the intermediate points are taken from the nearest points,
where result is available.

The Monte-Carlo temperature is gradually annealed from
$0.24 L/N$ to $0.008 L/N$ ($N=400$ is the total number of units) 
in 5000 sweeps through the entire system ($20\times 20\times 5000$ steps). The resulting 
layout is then examined and the phases visually identified.

\subsubsection{Perturbation theory}             
             
{\it Salt and Pepper} layout is relatively easy to study 
due to its uniformity, and can be solved exactly (Chklovskii and Koulakov, 2000).
If a layout does not deviate significantly from {\it Salt and Pepper},
i.e. the weak segregation case takes place,
it can also be solved exactly. This implies that the wire length can be written 
as an explicit functional of density distribution of the units. Such functional
was evaluated and optimized with respect to the density variations 
by Chklovskii and Koulakov (2000). 
The optimization shows that ODPs are formed for the values of 
parameter $|N_s - N_o|/N_s > 0.02$. However the simulated annealing 
method cannot distinguish weak segregated ODP from {\it Salt and Pepper}
for $|N_s - N_o|/N_s < 0.2$. 
There are two reasons for the failure of simulated annealing to do so:
\begin{itemize}

\item
Simulated annealing is performed at small but finite temperature
that destroys weakly segregated ODP.

\item
The units can be either completely right or left eye dominated.
This implies that OD can change only sharply in the described annealing version.
This is useful for obtaining the strongly segregated
phases, which occupy major part of the parameter space.
However, in the weak segregation limit the local OD changes smoothly.
Thus used version of simulated annealing performs poorly at $N_s\approx N_o$.

\end{itemize}

We therefore replace the simulated annealing results by those
from Chklovskii and Koulakov (2000) at small values of parameter $N_s-N_o$
(see the phase diagram below).

\subsubsection{Calculation of the ODP period.}
\label{exenum}                                   

To evaluate the period of ODP precisely, we first determine the phase
({\it Salt and Pepper}, {\it Stripes}, or {\it Patches}) for the given set of parameters $N_s/N_o$ 
and $f_L$, using methods described above. We then take a lattice containing a large number of
units, which exceeds sufficiently the lattice used in simulated annealing. This is possible because
the method of determining period described below is much less time consuming than simulated annealing.
We then arrange the two types of units on the lattice, using ODP determined by the simulated annealing, and
vary the period of the pattern to find the period producing the minimum of the wire length.
Below we describe the procedure for both {\it Stripes} and {\it Patches} in more details.

{\it i) Stripes} 

To find the optimum period for stripes we use an array containing 300 by 300 units.
This array includes three periods of the stripes, which run parallel to one of the sides of the region.
Each period therefore includes 100 units, containing $n_L$ left and $n_R$ right eye units, $n_L+n_R=100$. 
By varying $n_L$ we accomplish the change in the filling fraction of the left (ipsilateral) eye, according to the
formula: $f_L=n_L/\left( n_L + n_R \right)$. 
We consider a string of 100 units at the center of the
array, which is representative of all the units in the configuration. For each of the
central units the computer program establishes connections, according
to the connection rules. Most of the calculations are done for $N_s+N_o=300$.
We check that results change for different $N_s+N_o$ in a predictable fashion (see below, Results, subsection~\ref{sp}).
Stripes therefore have a fixed period in terms of number of units (100). To find the optimal spatial period of the 
stripes we vary the shape of each elementary cell in the 300 by 300 array. 
Thus, if the rectangular cell dimensions are $a_x$ perpendicular
and $a_y$ parallel to the stripes, we vary both $a_x$ and $a_y$, keeping the area of elementary cell
$a_xa_y=1$ constant. By doing so we do not change the density of units, but vary the spatial OD period,
according to the formula $\Lambda = 100a_x$. For each value of $a_x$ the cells are reconnected according to the 
connection rules. Special care is taken about exclusion of the boundary effects by making sure that none of the
units on the edges of the array is connected to. After the optimum period is found the 
period in terms of number of units is changed from 100 to another value, closer to the value of spatial period, 
to check for the absence of geometric artifacts, associated with distortions of elementary cells.
The change of the spatial period after this procedure is typically absent but in extreme cases does not exceed 3\%.

{\it ii) Patches} 

Since our results indicate that a triangular crystal of {\it Patches} is formed (see Fig.~\ref{phases}J),
we consider an array in the shape of parallelogram commensurate with the triangular arrangement of {\it Patches}.
The lattice sites in the array, representing units, are also arranged on a triangular lattice. Their positions are given by
$x(i,j)=i+j/2$ and $y(i,j) = j\sqrt{3}/2$, where $i$ and $j$ are integers varying between 1 and $5l$.
Here $l$ is the period of ODP to be optimized. 
The centers of {\it Patches} are located at points $x_c{n,m}=ln+lm/2$ and $y_c(n,m)=lm\sqrt{3}/2$.
Each patch includes lattice sites at the distance from a center determined by the filling fraction of the
ipsilateral eye: $R=l\sqrt{f_L\sqrt{3}/2\pi}$. The units within/outside the patch are left/right eye dominated.
The units in the configuration are then represented by the central parallelogram: $i,j=(2l+1)..3l$. 
For each of the units connections are made according to the connection rules with $N_s+N_o=300$.
The optimum period is obtained by varying parameter $l$.

\subsubsection{Fourier analysis of the ocular dominance patterns}             
\label{fourier}

\begin{figure}
\centerline{
\includegraphics[width=3.1in]{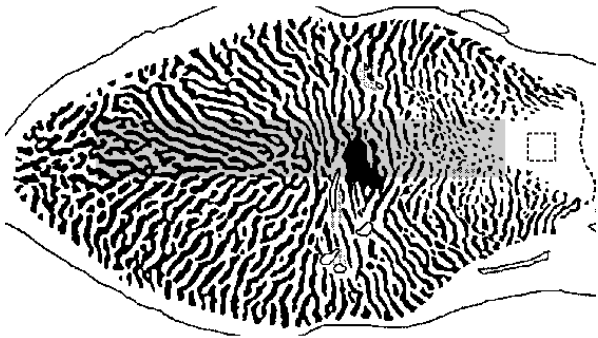}
} 
\caption{   
\protect{The image of the striate cortex of macaque monkey 1 left 
hemisphere from Horton and Hocking (1996a).
The left/right eye dominated areas are shown by black/white.
For the pixels in the shaded area we evaluate the filling factor
and OD period, displayed in Fig.~\ref{fp} below. The dashed square
gives an example of the region containing 64 by 64 pixels,
for which the filling fraction and Fourier transform are calculated.
It has dimensions 3.7 by 3.7 mm.
}
\label{monkey1}
}
\end{figure}

To determine the experimental dependence of the ODP period on the filling factor,
the image of ODP in macaque monkey (Horton and Hocking, 1996a) is converted into a digital format. 
In this format the image is represented by a set of pixels.
A pixel with coordinates $x$ and $y$ is represented by a number $s(x,y)$,
equal to $0$ for the right eye dominated and $1$ for the left eye dominated 
area. For each position in the image we then determine the local value of the average filling fraction
of the ipsilateral eye and the value of local OD period. Both these calculations are similar.
To do the calculation at a certain point in the map, given by coordinates $\left(x_0,y_0\right)$, 
we surround the corresponding pixel by a square, containing $64\times 64$ pixels 
(black square in Fig.~\ref{monkey1}, $3.7\times 3.7$mm).
The dimensions of the square are such that one hand it contains a few ODP periods (about 3),
which is needed for averaging, and on the other hand it is small compared to the 
characteristic dimensions over which the properties of ODP change ($\sim$ 1cm, see Fig.~\ref{monkey1}). 
To determine the filling fraction we average the scanned image over the square:
for position $\left(x_0,y_0\right)$ in the map the local value of the average filling fraction is given by
\begin{equation}
f_L \left(x_0,y_0 \right) = \frac {1}{64\times 64} \sum_{x=x_0-31}^{x_0+32}
\sum_{y=y_0-31}^{y_0+32} s \left(x,y\right).
\end{equation}

%
%

\begin{figure}
\centerline{
\includegraphics[width=3.1in]{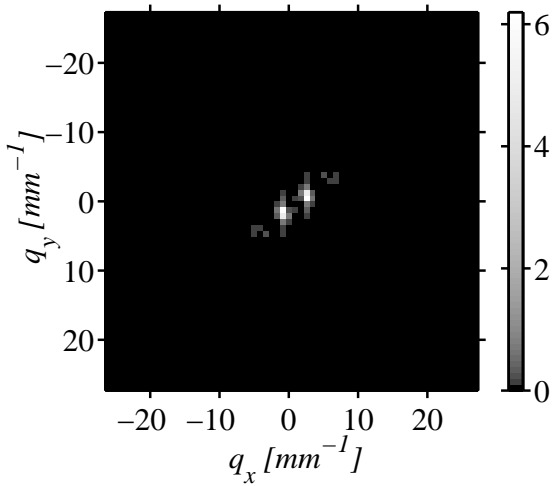}
} 
\caption{   
The spectral power for a point in the pattern occupied by stripes.
The spectrum has the bimodal appearance, characteristic of stripes.
The direction of the modes is determined by the direction in which ODP
changes (perpendicular to stripes). The distance of the modes from the
center determines the local value of ODP period by Eq.(\ref{odpp}).
The spectral power in the scale bar is in arbitrary units.
\label{tf}
}
\end{figure}

To determine the local value of ODP period we perform the Fourier transform of the
$s\left(x,y\right)-f_L$ in the square. As a result we obtain a set of numbers $\tilde{s} \left(q_x, q_y\right)$,
representing the Fourier transform amplitudes, defined on a $64\times 64$ set of wave 
vectors $\left(q_x, q_y\right)$. The spectral power, represented by $\left| \tilde{s} \left(q_x, q_y\right) \right|^2$, 
is shown in Fig.~\ref{tf} for one of the points in the pattern, corresponding to {\it Stripes}. 
It clearly has a bimodal appearance, indicating the average in the square direction of the stripes. 
We then determine the average value of the wave vector, using the formula:
\begin{equation}
\left< q  \left( x_0, y_0 \right)\right>= \frac{\sum_{q_x, q_y} \sqrt{q_x^2+q_y^2} \left| \tilde{s} \left(q_x, q_y\right) \right|^2}
{\sum_{q_x, q_y} \left| \tilde{s} \left(q_x, q_y\right) \right|^2}.
\end{equation}

%
%

The value of the mean ODP period is then defined as
\begin{equation}
\Lambda \left( x_0, y_0 \right)= \frac{2\pi}
{\left< q  \left( x_0, y_0 \right)\right>} .
\label{odpp}
\end{equation} 
This value for each pixel in the shaded area in Fig.~\ref{monkey1} is shown in Fig.~\ref{fp}.


\section{RESULTS}

\subsection{Small number of connections}
\label{qualitative}

We start by finding optimal layouts for three illustrative examples of 
wiring rules with small numbers of connections, $N_s$ and $N_o$. We 
caution the reader that because of the small numbers of connections
phase assignments may seem arbitrary. These examples are
chosen to illustrate our main results which will be confirmed both in
the lattice model with large $N_s$ and $N_o$ later in this section and in 
the continuous model (Chklovskii and Koulakov, 2001).

\begin{figure}
\centerline{
\includegraphics[width=3.1in,bbllx=100pt,bblly=338pt,bburx=464pt,bbury=491pt]{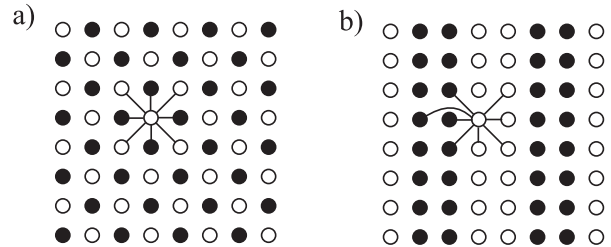}
}
\caption{ Ocular dominance patterns for $f_L=1/2$ and
$N_s=N_o=4$.  (a) A realization of the {\em Salt and Pepper} phase
gives minimal wire length (\protect{$l\approx 9.67$} lattice constants per neuron).  
(b) A realization of the {\em Stripe} phase is suboptimal (\protect{$l\approx 10.24$}).
\label{lat1}
}
\end{figure}

For the first two examples we set equal numbers of left and right
dominated neurons, $f_L=f_R=1/2$.  In the first example each neuron connects
with equal numbers of the same-eye and other-eye neurons,
$N_s=N_o=4$. Then the optimal layout is the ``chess board''  
of left/right neurons, Fig.\ref{lat1}a. This layout is a realization
of the {\em Salt and Pepper} phase, Fig.\ref{phases}a, because each
neuron has an equal number of left and right neurons among its immediate
neighbors. To calculate the length of connections per neuron, $l$, we
notice that in this layout all neurons have the same pattern of
connections. By considering one of them, Fig.\ref{lat1}a, we find that
$l=4+4\sqrt{2}\approx 9.67$. This layout is optimal because each neuron makes all of its connections 
with immediate neighbors.

%
%

A suboptimal layout for the same wiring rules is illustrated by a
realization of the {\em Stripe} phase, Fig.\ref{lat1}b. In this layout
each neuron has the same pattern of connections up to a mirror
reflection. By considering one of them, Fig.\ref{lat1}b, we find
$l=6+3\sqrt{2}\approx 10.24$, greater than $l\approx 9.67$ for the {\em
Salt and Pepper} phase. Here each neuron has among its immediate
neighbors only three other-eye neurons, while the wiring rules require
connecting with four other-eye neurons. A connection to a more distant 
neighbor is longer making the layout suboptimal. We confirm
the optimality of the {\em Salt and Pepper} phase for $N_s=N_o$ for large $N_s$, $N_o$ both
numerically and analytically.

%
%
\begin{figure}
\centerline{
\includegraphics[width=3.1in,bbllx=100pt,bblly=338pt,bburx=464pt,bbury=491pt]{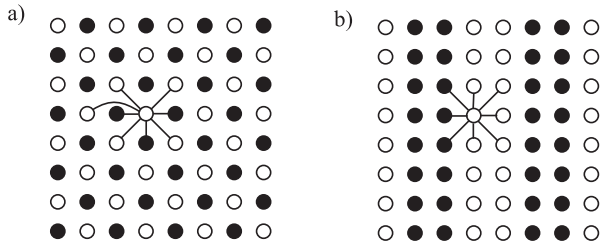}
}
\caption{ Ocular dominance patterns for $f_L=1/2$ and
$N_s=5$, $N_o=3$.  (a) A realization of the {\em Salt and Pepper} is
suboptimal (\protect{$l\approx 10.24$}).  (b) A realization of the {\em Stripe} phase gives minimal
wire length (\protect{$l\approx 9.67$}).
\label{lat2}
}
\end{figure}

In the second example each neuron connects with more same-eye than
other-eye neurons: $N_s=5$, $N_o=3$. Then a realization of the {\em
Salt and Pepper} phase, Fig.\ref{lat2}a is not optimal anymore. The
length of connections per neuron is $l\approx 10.24$, while the {\em
Stripe} phase, Fig.\ref{lat2}b gives $l\approx 9.67$. The {\em Salt and Pepper}
phase loses in wiring efficiency because there are not enough same-eye
neurons among immediate neighbors and connections with more distant
neighbors are needed.  The {\em Stripe} phase, Fig.\ref{lat2}b
rectifies this inefficiency by having each neuron make connections
only with immediate neighbors. Thus, clustering of same-eye neurons is
advantageous if each neuron connects more with the same-eye than with
the other-eye neurons.

%
%
\begin{figure}
\centerline{
\includegraphics[width=3.1in,bbllx=100pt,bblly=138pt,bburx=464pt,bbury=491pt]{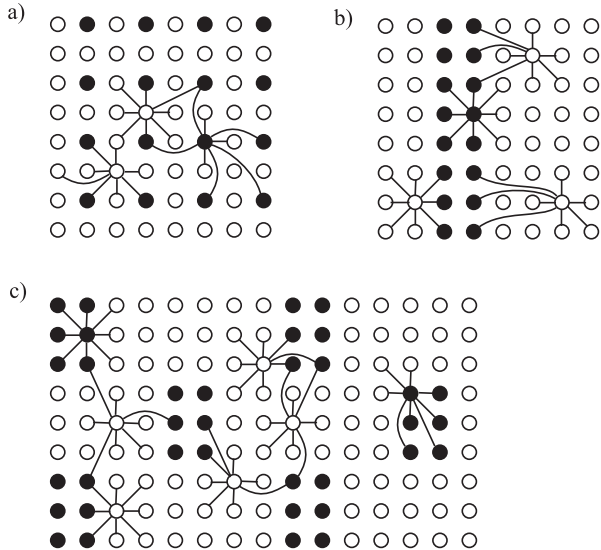}
}
\caption{ Ocular dominance patterns for $f_L=1/4$ and
$N_s=5$, $N_o=3$.  Realizations of the (a) {\em Salt and Pepper} (\protect{$l\approx 11.26$}) and
(b) {\em Stripes} (\protect{$l\approx 11.49$}) are suboptimal.  (c) A realization of the {\em
L-Patch} phase gives minimal wire length (\protect{$l\approx 10.67$}).
\label{lat3}
}
\end{figure}

In the third example we use the same wiring rules ($N_s=5$, $N_o=3$)
but take different numbers of left/right neurons, $f_L=1/4$,
$f_R=3/4$. The realizations of the {\em Salt and Pepper} phase is
shown in Fig.\ref{lat3}a and of the {\em Stripe} phase in
Fig.\ref{lat3}b. In these layouts, different neurons have different
patterns of connections. To find the wiring length per neuron we
average over different patterns and find for the {\em Salt and Pepper}
phase $l\approx 11.26$ and for the {\em Stripe} phase $l\approx 11.49$. A more
efficient layout is the {\em L-Patch} phase, Fig.\ref{lat3}c, where
$l\approx 10.67$. Although we cannot prove that the {\em L-Patch}
phase is optimal, this seems likely. Thus, the optimal shape of
monocular regions depends on the relative numbers of left/right
neurons.


\subsection{The shape of OD columns}
\label{PhDsection}

\begin{figure}
\centerline{
\includegraphics[width=2.0in,bbllx=97pt,bblly=224pt,bburx=250pt,bbury=598pt]{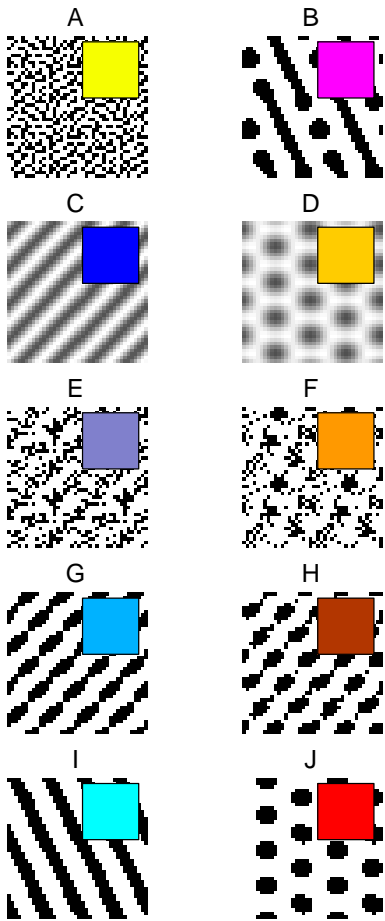}
}
\caption{
The phases obtained by perturbation theory ({\bf C} and {\bf D}) and by simulated annealing (the rest).
Each simulation array of 20 by 20 units in reproduced 4 times in each Figure.
{\protect The color bar is the key for Figure~\ref{PhD}}.
{\bf A}: {\it Salt and Pepper}; {\bf B}: {\it Stripes} mixed with {\it Patches};  
{\bf C} and {\bf D}: weakly segregated {\it Stripes} and {\it Patches} obtained by the perturbation theory; 
{\bf E} and {\bf F}: weakly segregated {\it Stripes} and {\it Patches} obtained by simulated annealing;
{\bf G}: modulated {\it Stripes}; 
{\bf H}: elongated {\it Patches}; 
{\bf I}: sharp {\it Stripes}; 
{\bf J}: sharp {\it Patches}; 
\label{phases}
}
\end{figure}

After giving some examples of ODPs with small numbers of connections $N_s$ and $N_o$
we discuss the opposite case of large numbers. As we show below in Section \ref{sp}, 
the shape of OD columns in this case does not depend on the absolute values of parameters $N_s$ and $N_o$. 
It is determined by the ratio $N_s/N_o$ and by the relative amount of ipsilateral neurons $f_L$ 
(assuming that the left eye is ipsilateral).
Depending on the values of parameters $N_s/N_o$, and $f_L$, optimal layout
belongs to the one of the eight phases shown in Fig.~\ref{phases}, where
ipsilateral and contralateral-eye dominated neurons are shown by 
black and white regions respectively.           
These phases can be divided into three major classes. 
The first class is represented by the unsegregated {\it Salt and Pepper} layout, 
in which two types of neurons are uniformly intermixed (Figure~\ref{phases}A).
The second class includes 
{\it Stripy} layouts, shown in Figures~\ref{phases}C, E, G, I.
The third class consists of {\em Patchy} layouts, displayed in Figures~\ref{phases}D, F, H, G.  

%
%

We distinguish several subclasses of {\it Stripy} phases. First, it is the sharp {\it Stripes} (Figure~\ref{phases}I),
which consists of alternating lamellar monocular regions. 
Second, it is the weakly segregated {\it Stripes} (Figure~\ref{phases}C,E).
In this ODP the variation of density of left/right eye dominated neurons is small.
This is an intermediate pattern between the unsegregated {\it Salt an Pepper} and the sharp {\it Stripe} layouts.
This phase is therefore fragile and difficult to obtain numerically. 
In some cases simulated annealing can produce such a phase, Figure~\ref{phases}E. 
In the other cases the weak segregated phase 
can only be obtained by the perturbation theory, which can carefully account for 
a weak variation of neuronal density. Such case is shown in Figure~\ref{phases}C.
Third, we also obtain {\em Stripy} phases that show a tendency to become 
{\em Patches}, by e.g. their longitudinal modulation, such as shown in Figure~\ref{phases}G. 

\begin{figure}
\centerline{
\includegraphics[width=3.1in]{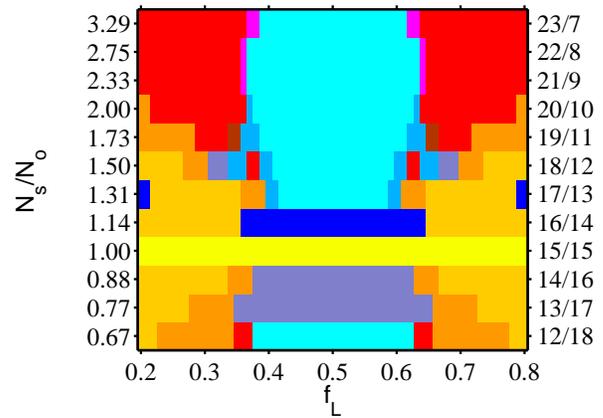}
} 
\caption{ The phase diagram showing the optimum phase for given values of parameters {\protect $f_L$ and $N_s/N_o$}.
For the color key see {\protect Figure~\ref{phases}}.
\label{PhD}
}
\end{figure}

Similar subclasses exist among {\it Patchy} layouts. 
We obtain sharp, weakly segregated (obtained from simulated annealing or perturbation theory), 
and elongated {\it Patches}, which are shown in Figures~\ref{phases}J, F, D, and H respectively.
Finally, there are mixed phases containing both {\it Stripes} and {\em Patches}, such as in Figure~\ref{phases}B.
These ODP's are shown on the phase diagram (PD) in Figure~\ref{PhD}.
The phase diagram shows the optimum phase (minimizing the total wire length)
for given values of parameters $N_s/N_o$ and $f_L$. 

The important feature of the PD is its left-right eye symmetry. 
It is apparent from the symmetry of Figure~\ref{PhD} with respect to the line $f_L=1/2$.
This is a consequence of the left-right eye symmetry of our model, implying that
the connection rules, defined by numbers $N_s$ and $N_o$ are independent on 
whether a neuron is left or right-eye dominated. For this reason the 
phase for $f_L>1/2$ can be obtained from the point with the same $N_s/N_o$ and the
value of the filling fraction equal to $1-f_L < 1/2$. This corresponds to the 
replacement of the white regions in Figure~\ref{phases} by black and vice versa.
A similar correspondence takes place when one compares ODP's in left and right hemisphere.

%
%

Another important feature of the PD is the existence of the {\it Salt and Pepper} region around 
the line $N_s/N_o=1$. This implies that the difference between $N_s$ and $N_o$ is the driving force of segregation into ODP. 
The larger the difference, the sharper the ODP becomes. 

\begin{figure}
\centerline{
\includegraphics[width=3.1in,bbllx=94pt,bblly=366pt,bburx=350pt,bbury=454pt]{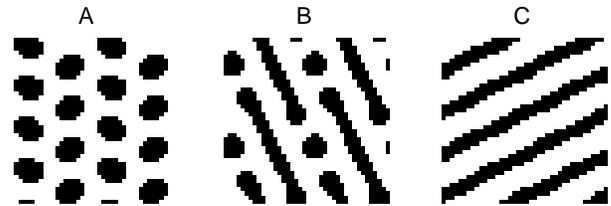}
} 
\caption{ The example of transition from {\it Patches} to {\it Stripes} at \protect{$N_s/N_o = 3$}.
{\bf A}, {\bf B}, and {\bf C} show the optimum phases for \protect{$f_L=0.34$, $0.36$, and $0.38$}
respectively.
\label{transition}
}
\end{figure}

The area of the PD adjacent to $f_L=f_R=1/2$ is occupied by {\it Stripy} ODPs. At small values of the filling fraction
the phases are {\it Patchy}. A transition from {\it Stripes} to {\it Patches} occurs at $f_L \approx 0.38$ almost
independently on parameter $N_s/N_o$. An example of such transition for $N_s/N_o = 3$ is shown in Figure~\ref{transition}.

%
%

\begin{figure}
\centerline{
\includegraphics[width=3.1in,bbllx=137pt,bblly=226pt,bburx=490pt,bbury=581pt]{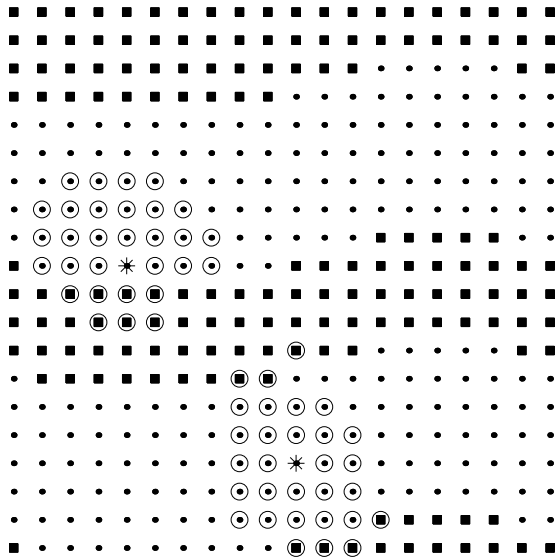}
} 
\caption{   
Connection arbors of two units in the array: near the interface of OD columns (top) and in the center of a column (bottom).
The units dominated by left and right eyes are shown by squares and dots respectively.
Two units for which the arbors are displayed are shown by stars. 
They are connected to other units, which are encircled.
The parameters for this particular layout are: \protect{$N_s=23$, $N_o=7$, $f_L=0.40$}. 
The connection arbor of the boundary unit (top) to the units of the same OD is significantly deformed,
compared to the corresponding connection arbor of the unit in the middle of the column (bottom).
This is similar to the observation of Katz et al., (1989). 
This deformation gives rise to the surface cost associated with formation of the interface
between columns. 
\label{connections}
}
\end{figure}

The reasons for the transition for {\em small} values of $N_s/N_o-1$ are discussed in  
Koulakov and Chklovskii, 1999. For {\em larger}differences, when ODP becomes sharp,
the transition occurs due to the presence of surface contribution to the 
wire length. To demonstrate this we present the following argument, which is rigorously valid in the asymptotic limit of large 
number of connections to the same-eye neurons, i.e. $N_s >> N_o$. 
In this limit connections to the same-eye neurons are the most abundant and therefore most costly, from wire length prospective.
Hence, we can disregard connections to the opposite-eye neurons in the first approximation.
Consider a unit near the interface between two OD columns (top unit in Fig.~\ref{connections}).
The connection arbor of this unit to the same OD units, shown by empty circles in Fig.~\ref{connections}, is strongly biased
toward the home column, since the unit has to maintain certain number of connections there.
This effect has been seen by Katz et al., 1989, in macaque striate cortex (see also the discussion in the Model Section above). 
For units away from the interface the connection arbor within the same OD column is close to a circle (Fig.~\ref{connections} bottom unit).
Thus, circular arbor renders the minimum wirelength in the absence of constraints, such as the interface between OD columns.
With the interface present the connection arbor to the same eye neurons is therefore strongly deformed
with respect to the optimum. Hence, the presence of the interface between the OD columns implies an increase
in the wirelength, and is therefore associated with a surface cost (similar to Malsburg, 1979). This surface cost drives
the transition from {\it Stripes} to {\it Patches}. Indeed if
$f_L \ll 1/2$ the patchy phases have much shorter length of the surface compared to {\it Stripes}. 
This is because {\it Patches} shrink when $f_L\rightarrow 0$ reducing their surface length, 
whereas {\it Stripes} become narrower, keeping their surface length the same. 
However, this is not true for $f_L=f_R=1/2$ where {\it Stripes}
have a shorter surface for numerical reasons. Therefore, when $f_L$ is decreased, the 
{\it Stripes} should unavoidably condense into {\it Patches} to minimize the surface cost.
This is similar to droplets of water assuming circular shape to minimize the surface energy.

%
%

We conclude therefore that two important transitions occur in our model.
\begin{itemize}
\item The transition from unsegregated {\it Salt and Pepper} to weakly segregated and then
sharp ODP is driven by the difference between parameters $N_s$ and $N_o$ characterizing the intra-cortical circuitry.  
\item The transition from {\it Stripy} to {\it Patchy} ODP is driven by the decreasing filling fraction
of the ipsilateral eye and occurs at $f_L \approx 0.4$, if left eye is underrepresented.
\end{itemize}

\subsection{Transition from {\it Stripes} to {\it Patches}: comparison to experiments}

Our phase diagram in Fig.~\ref{PhD} shows
that the transition from {\it Stripes} to {\it Patches} occurs when $f_L\approx 0.4$ for a wide range of $N_s/N_o$. 
This value will be compared now with the value of $f_L$ at which the transition occurs in the experiment, 
found from the relative area occupied by left eye dominated neurons. 
The conclusion that the {\em Patch} phase becomes optimal when contralateral eye dominates is, indeed, non-trivial,
because there may be a system of alternating wide and narrow monocular stripes instead.

\begin{figure}
\centerline{
\includegraphics[width=3.1in]{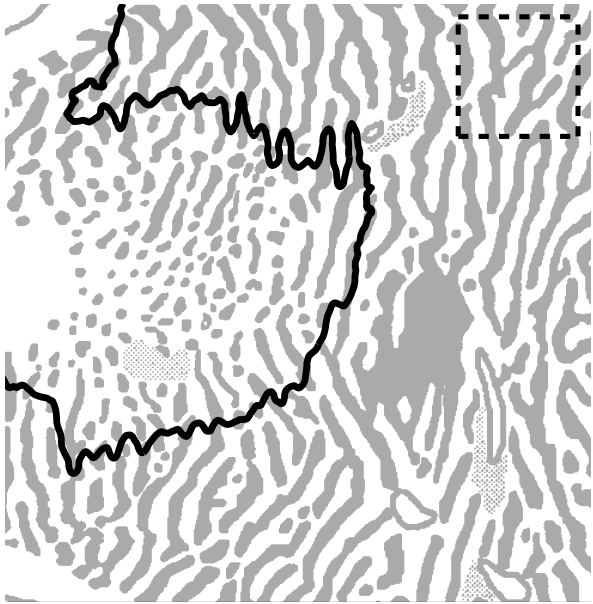}
}
\caption{Transition between the {\it Stripes}
and {\it Patches} occurs at theoretically predicted value
$f_L$. Shown is a fragment of the macaque ODP
from Horton and Hocking (1996a). Areas dominated by the
left eye are grey and neurons dominated by the left eye are white. 
The black contour corresponds to {\protect $f_L=0.4$} averaged over a window, whose dimensions are shown by the dashed square 
({\protect $3.7\times 3.7$}mm). 
The points of transition from {\em Stripes} to {\em Patches} coincides with the black contour.
\label{macaque}
}
\end{figure}

We test our conclusion on the data from macaque monkey first (Horton and Hocking, 1996a). 
The relative area occupied by the left/right eye depends on the location in V1. 
In the parafoveal part of V1 both eyes are represented equally, i.e. $f_L\approx 0.5$. 
ODP has a stripy appearance, in agreement with the phase diagram. 
Away from the foveal region contralateral eye becomes dominant. 
The ODP becomes patchy there (LeVay et. al., 1985), just as
expected from the theoretical phase diagram. We verify the location of the
transition by using the following algorithm. We find $f_L$ for each
point of the pattern by calculating the relative area occupied by the
left/right regions in a window centered on that point and a few OD
periods wide (dashed lines in Fig.~\ref{macaque}). Then we draw a contour 
corresponding to $f_L=0.4$, Fig.~\ref{macaque}.  We observe in  Fig.~\ref{macaque} that
stripes indeed become patchy at the black contour indicating $f_R=0.4$.

%
%

In {\em Cebus} monkey the ODP has a similar transition (Rosa et al., 1992).
For monkey CO6L from Rosa et al., 1992, we determine visually that 
along the horizontal meridian the transition occurs at the eccentricity of 
$20-40\deg$. According to the plot of the relative representations given
in  Rosa et al., 1992, $f_L$ changes in the range $0.32-0.42$ at these
eccentricities. Our theoretical conclusion about a transition at $f_L=0.4$ falls into this interval. For 
the upper $45$ degree meridian of the same monkey
the transition occurs at the eccentricity of $30-40$ degrees or at filling
fractions $0.33-0.43$. Again, the predicted value belongs to this
interval. We conclude that these data are consistent with the results of our model.

In cats the ODPs have a patchy appearance (Anderson et al., 1988; Jones et al., 1991). 
In this case our theory implies that one of the eyes should dominate. According to some reports 
(Shatz and Stryker, 1978; Crier et al., 1998) the filling fraction of the
contralateral eye in cat V1 is about $0.8$ in young animals (before P22). 
This may lead to {\em Patches} in cat V1. 
The strong contralateral bias disappears in older animals (Crier et al., 1998).
This is consistent with other reports (Anderson et al., 1988) 
that both eyes are represented almost equally in adult cats.

\subsection{The period of ocular dominance pattern}

\subsubsection{Scalability of the Model}
\label{sp}

One of the general features of our model is that the period of OD pattern becomes larger, 
when the total number of connections is increased. Indeed, the size of the connection arbors grows 
if both $N_s$ and $N_o$ are increased, given that the density of units ($1/(30\mu\times 30\mu)$) is kept constant. 
This is because one has to go further to find the necessary number of connections to satisfy the wiring rules.  
Since dimensions of the connection arbors set up a characteristic scale for the 
OD pattern, the period of the latter also increases. This property of our model, 
which we call {\it scalability}, is discussed in this subsection.

\begin{figure}
\centerline{
\includegraphics[width=3.1in,bbllx=90pt,bblly=377pt,bburx=531pt,bbury=597pt]{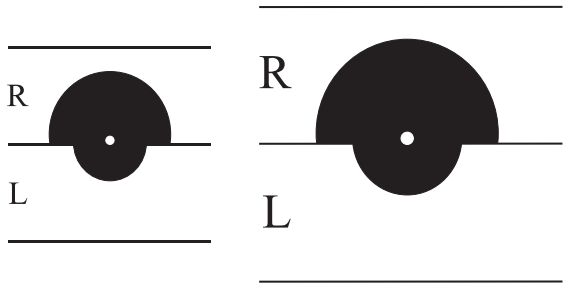}
} 
\vspace{0.5in}
\caption{   
The original ODP and connection arbor for one of the neurons, marked by the white dot (left).
Compare to Fig.~\ref{connections}, top neuron. The stretched configuration is shown on the right.
\label{scalaba}
}
\end{figure}

Let us define scalability in a mathematically rigorous fashion.
Assume that one has found a minimum wire length configuration for certain set of parameters $N_s$, $N_o$, and $f_L$ (or $f_R=1-f_L$).
Assume that both $N_s$ and $N_o$ are very large. In this case the dimensions of connection arbors are much larger than the
lattice spacing, and one can ignore the fine structure imposed on the connection arbors by the square lattice.
This is exactly the limit in which our model has some validity, both because realistic numbers of neuronal 
connections are large and because we would like to avoid artifacts introduced by the square lattice.
An example of connection arbors for a neuron for some set of $N_s$ and $N_o$ are shown in Fig.~\ref{scalaba} (left).
This neuron and its connection arbors resemble the top neuron, marked by the star, in Fig.~\ref{connections}.
The connection arbors in Fig.~\ref{scalaba} look like continuous circular pieces, 
due to the large $N_s$ and $N_o$ limit (square lattice makes the boundaries of connection arbors look like staircases,
whose steps are too small to show in the picture for large $N_s$ and $N_o$). 
Imagine now a geometric transformation, in which the dimensions of the
connection arbors of all of the neurons, as well as the OD pattern itself, are blown up by the same scaling factor $\eta > 1$.
The two-dimensional density of the neurons must be preserved during this transformation.
The obtained new OD pattern and the new connection arbors are shown schematically in Fig.~\ref{scalaba} (right). 
Since the transformation is applied to the two-dimensional objects, and each of the dimensions was 
stretched by the factor $\eta$, each neurons in the new configuration will receive $\eta^2N_s$ and $\eta^2N_o$
connections from the same and opposite eye neurons. The newly obtained configuration (Fig.~\ref{scalaba} right)
will satisfy wiring rules with connection numbers given by $\eta^2N_s$ and $\eta^2N_o$. 
Note that the filling fraction is not changed by this transformation.
It is $f_L=f_R=1/2$ in Fig.~\ref{scalaba}. Will this be the minimum wire length
configuration for this set of parameters?

To prove that the new configuration minimizes the total wire length for the new set of parameters
$\eta^2N_s$ and $\eta^2N_o$ we notice that the total wirelength for the new configuration
is given by $\eta^3 L$, where $L$ is the total wirelength for the original configuration (Fig.~\ref{scalaba} left).
This is because each neuron now receives the number of connections increased by $\eta^2$,
and each connection is stretched by $\eta$. Imagine now that one finds a new configuration,
which has the total connection length $L'<\eta^3 L$. Let us take this more optimal configuration
and shrink it by the factor of $\eta$. We obtain a configuration, satisfying wiring rules for the set $N_s$ and $N_o$,
whose total wirelength is $L'/\eta^3 < L$. But this contradicts to our postulate that the original configuration
in Fig.~\ref{scalaba} (left) is optimal for the set of parameters $N_s$ and $N_o$. Thus the stretched configuration 
provides the minimum of the wirelength for the new set of parameters $\eta^2N_s$ and $\eta^2N_o$. 

This property is important, because once the solution for given $N_s$ and $N_o$ is found, one can obtain 
other solutions, with the set of parameters $\eta^2N_s$ and $\eta^2N_o$, 
by stretching the original configuration by the factor of $\eta$ uniformly in all the directions. 
The important property which remains the same for all these related configurations is the ratio between the numbers of the same and other
eye connections, $N_s/N_o$. Thus, we conclude that this ratio determines the shapes of the OD patterns,
which is unchanged during the uniform stretching procedure.  

What is changed in the uniform stretching is the ODP period?
Since the period is proportional to the stretching parameter $\eta$, while the total
number of connections is proportional to $\eta^2$, we conclude that the period is proportional 
to the square root of the total number of connections, when the
ratio $N_s/N_o$ is kept constant. This implies that
\begin{equation}
\Lambda \left( N_s, N_o, f_L \right) =   D \cdot \lambda \left( N_s/N_o, f_L\right)
\label{scalab_form}
\end{equation}
Here $D = 2 a \sqrt{\left( N_s + N_o \right)   / \pi} \sim \eta$, where $a=30\mu$ is the size of the microcolumn unit.
Parameter $D$ has a meaning of the average axonal arbor diameter.
The quantity $\lambda(N_s/N_o, f_L)$ is the {\it normalized OD period}, calculated in the units of the average axonal diameter.
This quantity is introduced here for easier comparison to the experiment.
Notice that this quantity does not depend on the total number of connections.
The latter dependence is entirely absorbed by the parameter $D$.

\begin{figure}
\centerline{
\includegraphics[width=3.1in]{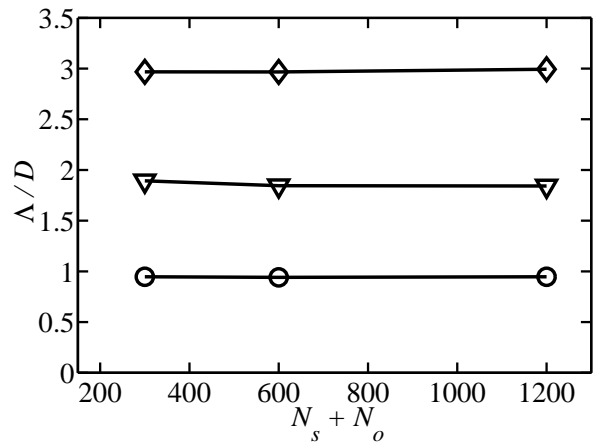}
} 
\caption{   
The independence of the ratio \protect{$\Lambda/D$ [see Eq.(\ref{scalab_form})]} on the 
total number of connections. This implies that the lattice effects, violating scalability,
are insignificant. The data are obtained for \protect{$f_L=1/2$}. Circles, triangles, and 
diamonds show results for \protect{$N_s/N_o$} equal to 4/3, 17/3, and 14 respectively.
\label{scalab}
}
\end{figure}

Scalability is valid for the limit of large $N_s$ and $N_o$, when square lattice effects can be ignored, and our model
becomes continuous. Does scalability apply to our numerical solution, described in subsection~\ref{exenum}?
To check this we plot the ratio $\Lambda \left( N_s, N_o, f_L \right)/D$, obtained using methods described in~\ref{exenum}, 
for different values of the total number of connections $N_s + N_o$ in Fig.~\ref{scalab}. 
If Eq.~(\ref{scalab_form}) is valid, this ratio should not depend on the total number of connections.
As evident from Fig.~\ref{scalab} this property is indeed satisfied. Hence, below in this subsection 
we always present the results for $\lambda \left( N_s/N_o, f_L\right)= \Lambda/D$, 
which are independent on the total number of connections,
assuming that the latter dependence can be easily recovered using Eq.~(\ref{scalab_form}).

%
%

\subsubsection{Dependence on parameter $N_s/N_o$}

\begin{figure}
\centerline{
\includegraphics[width=3.1in]{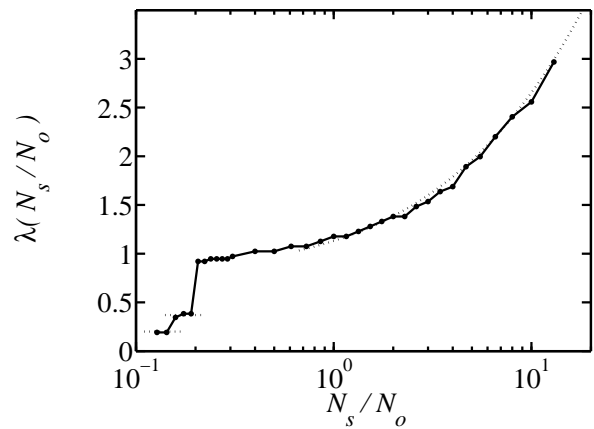}
} 
\caption{   
Dependence of the normalized OD period, defined in \protect{Eq.~(\ref{scalab_form})}, on the parameter \protect{$N_s/N_o$}.
The data points are shown by dots connected by lines. The dotted curve shows the asymptotic fit obtained for the values
of parameters \protect{$N_s/N_o >> 1$ [Eq.(\ref{asymp_p})]}. Two horizontal dotted lines show the
plateau values of the period separated by discontinuous transitions at \protect{$N_s/N_o \approx 1.15$ and $1.20$}.
\label{fs}
}
\end{figure}

We now examine the dependence of normalized period $\lambda(N_s/N_o,f_L)$ [see Eq.~(\ref{scalab_form})] on 
the parameter $N_s/N_o$, for $f_L=1/2$, when we have to consider the stripe phase, 
according to subsection~\ref{PhDsection}.
The results are shown in Fig.~\ref{fs}. These results have been obtained using methods described in 
subsection~\ref{exenum}. 
In general, the normalized period increases with increasing parameter $N_s/N_o$. 
This increase in the OD period can be understood considering the interplay between connections to the same and opposite eye units. 
Indeed, the presence of connections between the same eye units implies affinity between the same OD neurons.
An increase in the relative number of such connections ($N_s/N_o$) strengthens such affinity.
The OD columns provide a neighborhood rich with the same eye neurons; even more so, on average, for coarser domains.
Thus stronger affinity between the same eye neurons ($N_s/N_o$) leads to an increase in the size of OD domains. 
This effect is produced by wiring economy principle, since the latter provides a substrate for the affinity of connected neurons.

%
%

We now examine Fig.~\ref{fs} in more detail. The relative period diverges in the limit $N_s>>N_o$. 
The divergence can be described by the asymptotic formula
\begin{equation}
\lambda(N_s/N_o, f_L=1/2) \approx 0.8 \sqrt{\frac{N_s}{N_o}}
\label{asymp_p}
\end{equation}
shown in Fig.~\ref{fs} by the dotted curve. The divergence can be understood as follows.
Imagine that the neurons do not have to connect to the neurons of the opposite OD, i.e.
parameter $N_o=0$, $N_s\neq 0$, which corresponds to the extreme case $N_s>>N_o$. 
In this case the optimum wire length configuration consists of only two large domains, dominated 
by left and right eye neurons, occupying a half of V1 each. This is because of the notion of surface 
contribution, introduced in subsection~\ref{PhDsection}. To minimize this interface contribution
the system phase segregates into two large domains. 
Thus, in the case $N_o=0$ ODP has maximum possible period, spanning the entire V1. 
This explains the tendency of the period diverge in the limit $N_o\neq 0$ ($N_s/N_o=\infty$) in Fig.~\ref{fs}.
What happens if $N_o\neq 0$? Since the neurons now have to connect to the opposite eye 
neurons, phase segregated configuration (two large domains spanning the entire V1) is no longer optimum. 
The system introduces more interfaces between OD columns
to shorten distances between opposite eye neurons. More interfaces implies a reduction in the OD period. 
Thus, finite $N_s/N_o$ leads to the finite OD period. This is reflected by the 
asymptotic dependence (\ref{asymp_p}) and the dotted curve in Fig.~\ref{fs}.

An interesting phenomenon observed in Fig.~\ref{fs} is the presence of a few discontinuous changes
in the OD period. One such a change occurs at $N_s/N_o \approx 1.15$ and consists in an abrupt
increase in the OD period by a factor of about $2.3$. Another discontinuous transition occurs at $N_s/N_o \approx 1.20$
and the corresponding increase in the period is by a factor of $2$. Note that these transitions are truly discontinuous,
i.e. they occur at discrete points. They appear smooth in Fig.\ref{fs} due to a sparse sampling
(the real data points are shown by dots).  Note also that the quantity $D$ in Eq.~(\ref{scalab_form}) changes negligibly in the
interval between $N_s/N_o=1.1$ and $1.2$, which implies that both OD period $\Lambda$ and the normalized period $\lambda$ change approximately
by the same factor. Such discontinuous changes in the OD period in the weakly segregated regime, i.e. 
when the ODP is not well defined, may be responsible for the coarsening of ODP in strabismic squirrel monkeys 
(see Discussion for more details).

\subsubsection{Dependence on the filling fraction}

\begin{figure}
\centerline{
\includegraphics[width=3.1in]{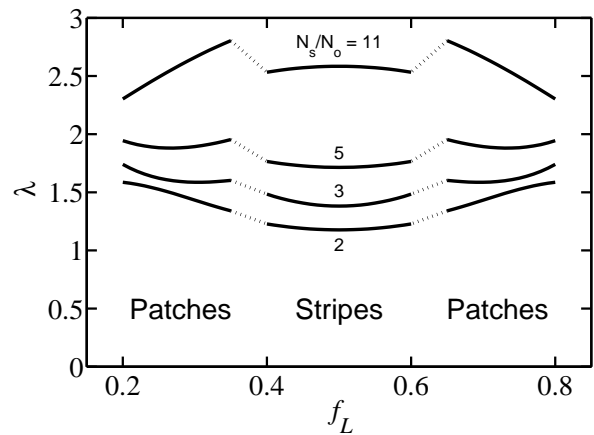}
} 
\caption{   
\protect{The normalized OD period $\lambda$ as a function of the filling fraction of one of the eyes.
Four dependencies are shown for the values of parameter $N_s/N_o$ equal to $2$, $3$, $5$, and $11$ correspondingly. 
The central segment of the dependencies, between $f_L=0.4$ and $0.6$ corresponds to the
{\it Stripe} phase. The dependencies in the regions between $0.2\leq f_L \leq 0.35$
and $0.65 \leq f_L \leq 0.8$ have been calculated for {\it Patches}, as indicated in the 
Figure. In the regions of transition between {\it Stripes} and {\it Pathes} the curves are 
connected by dotted lines.
}
\label{fr}
}
\end{figure}

The dependence of the normalized period $\lambda(N_s/N_o, f_L)$ on the filling fraction of the left eye $f_L$ is shown in
Fig.~\ref{fr}. These results have been obtained using methods described in 
subsection~\ref{exenum}. Four dependencies are shown, for four values of the parameter $N_s/N_o$ marked on each curve.
The general tendency for the period to grow with increasing parameter $N_s/N_o$, described in the previous
subsection, is evident in the Figure. 

%
%

For small values of parameter $N_s/N_o$ the period increases when the filling fraction moves away from
$f_L=1/2$, as long as one stays within the same phase ({\it Stripes} or {\it Patches}). 
This behavior is seen for the two bottom curves in Fig.~\ref{fr}.
In the transitional region between {\it Stripes} and {\it Patches} the OD period
experiences a discontinuity, marked by the dotted lines. For large values of $N_s/N_o$,
the dependence of the period on $f_L$ is opposite: the period decreases, as the filling factor
deviates from $1/2$, as demonstrated by the top curve in Fig.~\ref{fr}.

%
%

\begin{figure}
\centerline{
\includegraphics[width=2.5in]{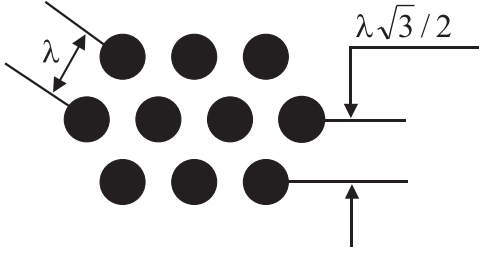}
} 
\caption{   
For {\it Patches} Fourier transform produces distance between rows, rather then the period of the OD pattern. 
The distance between rows is a natural successor of the period of {\it Stripes} after the transition
to {\it Patches} occurs (see also Fig.~\ref{transition}). The distinction between {\it Patch} period and
distance between rows should be taken into account for accurate comparison to the experimental observations.
\label{rows}
}
\end{figure}

\begin{figure}
\centerline{
\includegraphics[width=3.1in]{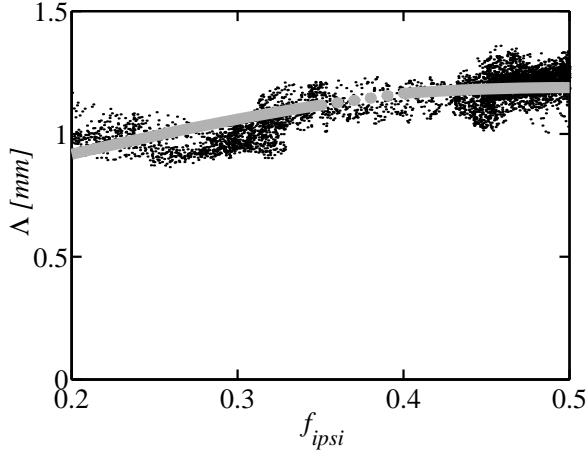}
} 
\caption{   
Comparison of the OD period observed in the experiment (Horton and Hocking, 1996a)
in macaque striate cortex (dots) to the results of our theory (gray curve).
The former is obtained using Fourier transform method described in subsection~\ref{fourier}.
The latter is the top curve in Fig.~\ref{fr}, with the sector of the data corresponding to {\it Patches}
corrected by the factor \protect{$\sqrt{3}/2\approx 0.87$} for compatibility with the Fourier transform.
The only fitting parameter used is $D=0.46$mm [see Eq.~(\ref{scalab_form})].
\label{fp}
}
\end{figure}

We now compare this behavior of our model to the observations in the striate cortex of macaque monkey (Horton and Hocking, 1996a),
using Fourier transform method described in subsection~\ref{fourier}. To make this comparison possible 
the following technical consideration is taken into account. The Fourier transform method 
evaluates the average value of the spatial frequency of the ODP $\left< Q \right>$.
The OD period is then calculated by the formula $\Lambda = 2\pi / \left< Q \right>$. 
For the {\it Stripe} phase this procedure results in the value, which is close to the 
period of stripes. For {\it Patches} it results in the distance between rows of patches,
which is smaller than the period by the factor $\sqrt{3}/2\approx 0.87$ (see Fig.~\ref{rows}). Thus, to make comparison
to the experiment possible, the values in Fig.~\ref{fr} corresponding to {\it Patches} should be multiplied by the
factor $0.87$. The result of this procedure is shown in Fig.~\ref{fp} by the gray line. 

%
%

Fig.~\ref{fp} shows that the period observed in the experiment decreases when 
the filling factor of the ipsilateral eye deviates from $1/2$. This warrants the use of 
the top curve in Fig.~\ref{fr} to represent the theoretical result. 
Since, the shape of the theoretical dependence does not change much
when $N_s/N_o > 10$, the parameter $N_s/N_o$ cannot be established from the
comparison of the theory to the experiment. 
To obtain the gray curve in Fig.~\ref{fp} the $N_s/N_o=11$
dependence in Fig.~\ref{fr} was multiplied by the fitting parameter 
$D=0.46$mm. This is the only fitting parameter used.
As seen in Fig.~\ref{fp}, our theory describes the experimentally
observed dependence quite well.

%
%

\begin{figure}
\centerline{
\includegraphics[width=3.1in]{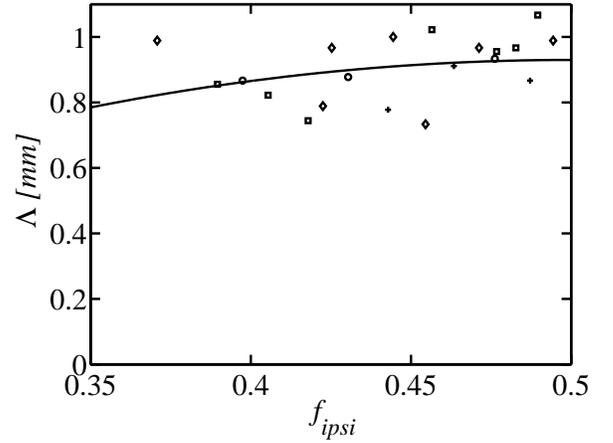}
}
\caption{  
The dependence of the ODP period on the filling fraction based on data from Tychsen and Burkhalter (1997).  
The results for non-strabismic adult macaque monkeys are presented by markers: 
monkeys M25 right V1 (diamonds), M25 left V1 (circles), 418 (squares), and 906 (dots).
The curve shows the best parabolic fit (see text). 
\label{tychsen}
}
\end{figure}

The widths of the ipsilateral and contralateral eye stripes in macaques has also been  
measured independently by Tychsen and Burkhalter (1997). 
Based on their results one can evaluate the ODP period and the filling fraction:
\begin {equation}
\Lambda = W_I+W_C,\ \ \ f_{ipsi} = W_I/\left( W_I+W_C \right).
\end{equation}
Here $f_{ipsi}$, $W_I$, and $W_C$ are the filling fraction of the ipsilateral eye, and the ipsilateral/contralateral 
column widths respectively. The dependence of the period on the filling fraction can therefore be established. 
This dependence is shown in Fig.\ref{tychsen}.

The best parabolic fit to the data in Fig.\ref{tychsen} is given by:
\begin{equation}
\Lambda(f)=\Lambda(1/2)\left[ 1-\alpha \left( f_{ipsi}-1/2 \right)^2\right].
\end{equation}
The coefficient $\alpha = -6.94\pm 6.38$ is estimated using bootstrap (Efron and Tibshirani, 1993). 
The expectation value of the coefficient is therefore below zero, as
seen from Fig.\ref{tychsen}. The probability of the coefficient to 
be greater than zero is $p=0.11$ as evaluated by bootstrap procedure, which is
used since the distribution of coefficients $\alpha$ is non-gaussian. 
This implies that with great degree of certainty one can assume that the period of ODP
does decrease with the filling fraction deviating from $1/2$. 

It should be noted that the value of coefficient $\alpha$ can be obscured by the variability of ODP period between different individuals,
since data in Fig.\ref{tychsen} are assembled from three monkeys (four V1's). To reduce the impact of inter-individual variability
we then normalized the period for each area V1 by the mean value for each individual animal. 
The value of the coefficient is then $\alpha=-7.98\pm 6.24$, with the probability
of positive coefficient $p=0.055$. Thus the decrease of the period with filling fraction is even more certain, when the inter-individual
variability is accounted for. The value of coefficient $\alpha$ obtained from the theory is $2$ (Fig.~\ref{fr}, $N_s/N_o=11$). 
It is consistent with both measurements.

%
%


\section{Discussion}
\label{Discussion}

In this work we propose a model which can account for most of experimentally observed features of ODPs. 
Our model has two principal parameters. The first parameter characterizes the intracortical circuitry.
It is the difference between the number of connections to the same and to the opposite OD neurons.
Our results suggest that this difference is the driving force of segregation into ODPs.
We argue therefore that the sensitivity of the intra-cortical connectivity to OD
provides a reason to formation of OD columns (see below). 
The second parameter is the fraction of neurons dominated by the ipsilateral eye.
This parameter determines the shape of monocular regions in ODP.
In the majority of the primary visual area of macaque and {\it Cebus} monkeys
this parameter is close to $50\%$, which implies that both ipsi- and contralateral
eyes are equally represented. However, in the proximity of monocular crescent
the ipsilateral eye becomes underrepresented. 
This is because the inputs into the eye are blocked by the nose of the animal.
Our theory suggests that the decrease in the filling fraction of the ipsilateral eye drives the transition in the ODP structure 
from stripy (zebra skin like) to patchy (similar to leopard skin). The transition occurs
when the fraction of the ipsilateral eye dominated neurons
approaches $40\%$ in both macaque and {\it Cebus} monkeys (see below).
We also analyze the dependence of OD period on the parameters of our model and find satisfactory agreement with 
experimental data.

\subsection{On the functional significance of OD columns}

Each neuron in our model establishes certain number of intra-cortical connections 
with neurons dominated by the same and the opposite eye. 
As suggested by experimental studies in macaque striate cortex, neurons in layer $4C\beta$
typically make more connections with neurons of the same OD (Katz et al., 1989).
Thus, from wiring economy prospective, connections with neurons of the same OD are more important than the opposite eye connections.
Therefore, it is advantageous to form OD columns, since they provide environment rich with the same OD neurons, 
which results in shortening connections to the same eye neurons. 
The wiring economy principle thus provides a natural reason for the existence of OD patterns, 
i.e. answers the first question in the program listed in the Introduction. 

Our model suggests that in primates with weakly defined OD columns, such as squirrel monkey (Hubel et al., 1976; Livingstone, 1996; 
Horton and Hocking, 1996b) and owl monkey (Livingstone, 1996), the difference between these two
components of intracortical connectivity is not large. Such difference may be
increased in these animals by experimentally induced strabismus. 
Indeed, strabismus reduces correlated activity between opposite OD cortical neurons, therefore
reducing their connectivity $N_o$. Reduction in $N_o$ unbalanced by the corresponding reduction in $N_s$
increases the parameter $N_s/N_o$
and leads to sharpening of OD columns, according to our phase diagram in Fig.~\ref{PhD}.
Such sharpening is indeed observed experimentally (Shatz et al., 1977; Lowel, 1994; Livingstone, 1996).
This phenomenon was also predicted theoretically by Goodhill (1993).

The two parameters of intra-cortical circuitry, $N_s$ and $N_o$, represent in our model the 
interplay between two classes of processing performed by the visual cortex. 
The first class includes the processing of the monocular image, for which connections to the same OD
neurons are necessary. The second class includes various tasks related to stereopsis,
which require comparison of two monocular images, relying on the connections between the opposite OD 
neurons. We proposed above that the function of OD columns is to shorten the connections between the same eye neurons. 
The impact of stereoscopic vision should therefore be the opposite: strong stereoscopy should make ODP weaker.
Inversely, weak stereoscopy induces sharp ODPs. The latter statement is justified by
the observations in animals with experimental strabismus. However, one should be careful about this statement, 
since the relation between functional and anatomical properties may not be direct.

The situation in the animals with lateral eye positioning, such as mice, rats, tree shrews, etc., is different. 
In such animals the visual pathway is almost completely crossed, i.e. V1 in each hemisphere
is strongly dominated by the contralateral eye [Drager, 1974, 1975, 1978; Drager and Olsen, 1980; Gordon and Stryker, 1996 (mouse); 
Hubel, 1977 (rat); Casagrande and Harting, 1975; Mully and Fitzpatrick, 1992 (tree shrew); Horton and Hocking, 1996b (other species)]. 
As suggested by Antonini et al. (1999) this implies that the ODP contains only two large monocular 'columns', each spanning the whole hemisphere. 
This can be interpreted as an OD having very large period, spanning both striate cortices.
This picture can be fitted into the framework of our model. 
Indeed, we predict that if the number of connections to the other OD neurons ($N_o$) is very small
the OD has a very large period (Fig.\ref{fs}). Thus our model predicts that the number of connections received by each neuron
from the neurons of the same OD ($N_s$) is much larger than the number of opposite eye connections ($N_o$)
in the species with lateral eye positioning. This should include the cross-hemispheric projections.
This statement should have functional consequences. Since $N_o$ is small, 
synthesis of images from two eyes is weaker. 
Therefore the animals have to find another strategy to implement stereopsis.
Hooded rats for example use vertical head movements to estimate distances (Legg and Lambert, 1990).
Our conclusion about small $N_o$ should also apply to the superior colliculus in these 
animals (Colonnese and Constantine-Paton, 2001), 
as well as to the tectum in lower vertebrates (Schmidt and Tieman, 1985), in which cases the visual inputs cross over almost completely too.

To summarize, our model encompasses most of the phenomena related to the 
sharpness and observability of ODP. It relates the interspecies variability
in the ODP to the relative amount of binocular interaction occurring in different species.
Thus, we predict, that in the animals with weakly segregated columns (squirrel monkey) $N_s\approx N_o$;
in the animals with sharp columns (macaque) $N_s$ is much larger than $N_o$ [confirmed by Katz et al., (1989)]; 
and, finally, in the animals with lateral eye positioning, $N_o$ should have a value, 
whose contribution to the wirelength can be neglected.

\subsection{Variation of the ODP period in the extrafoveal region}

Another consequence of a decreasing $f_L$ in macaque is a decrease in the ODP period (LeVay et al., 1985).
In Fig.~\ref{fp} we compare the result of our theory to the data from macaque monkey 
(Fourier transform applied to data from Horton and Hocking, 1996a).
From this comparison we conclude that, according to the wiring economy principle, parameter  $N_s/N_o>>1$, or cells establish much 
more connection with the same OD cells, than with the opposite. This result of 
is consistent with the observations of OD 
sensitive circuitry in the striate cortex of macaque by Katz et al. (1989). 

We chose the regions proximal to the horizontal meridian for this comparison. 
This is based on the assumption that cortical properties, such as $N_s$ and $N_o$, 
change little along this meridian. This assumption is in part supported by the fact
that OD periodicity changes little on the large segment of the meridian
occupied by stripes, spanning the region between about 2 and 25 degree eccentricity (notice very little scatter
in Fig.~\ref{fp} around the point $f_L=1/2$). The changes in the period begin to occur when
$f_L$ deviates from $1/2$. This is illustrated by Fig.~\ref{fp}. Other authors notice 
a decrease in the period when comparing vertical to horizontal meridian.
Studies based of computer reconstructions report about 2 fold decrease in OD periodicity
comparing these areas (LeVay et. al., 1985), while others, based of flat-mounts (Horton and Hocking, 1996a),
report a more moderate change. Such variation cannot be accounted for by a decrease in parameter 
$f_L$ alone, since $f_L$ is about $1/2$ on both meridians in close proximity to parafoveal region (<20 degrees
of eccentricity). Our model suggests two possibilities based on the variation in the 
intracortical circuitry, described by $N_s$ and $N_o$. Since such differences 
in the circuitry may also be responsible for variability of the OD period 
between different animals, we discuss this possibility in the next subsection.

\subsection{Variability of the ODP period from individual to individual}

Studies in macaque monkeys (Horton and Hocking, 1996a) reveal large
inter-individual variability of the stripe period. The stripe period
is $1072\pm 164\mu$ along the V1 border, after comparison of 6 animals. 
Two factors may contribute to this phenomenon in the framework of our model.
(1) The basic diameter of axonal and dendritic arbors $D$ varies from animal to animal.
This could be due to changes in $N_s$, $N_o$, or neuronal density.
(2) The ratio between monocular and binocular interactions $N_s/N_o$ varies.
The former reason is justified by Eq.~(\ref{scalab_form}). The latter can be understood from
Fig.~\ref{fs}. Simply speaking, monocular interactions ($N_s$) favor formation of OD columns,
making them wider, in an effort to provide same OD rich environment for all the neurons. 
Binocular interactions ($N_o$) favor interfaces between columns, since interfaces bring
opposite OD neurons closer to each other. They therefore decrease OD period.
When $N_s/N_o$ increases the OD period increases too (Fig.~\ref{fs}). 
This may occur when comparing different individuals.

\subsection{ODP period in strabismic animals}

Similar consideration may apply to the experiments in strabismic animals (Lowel, 1994; Livingstone, 1996).
Since strabismus reduces correlations between eyes, its effect in our model 
is to reduce parameter $N_o$. Hence, the ratio $N_s/N_o$ is increased.
According to our model (Fig.~\ref{fs}) this generally leads to an increase in the relative OD period 
(ratio of the basic OD periodicity to the connection range $D$). This result is understood
from the interplay between affinity between the same eye neurons ($N_s$), increasing the period, 
and the affinity between opposite OD neurons ($N_o$), reducing OD period. 
Since the latter is reduced by strabismus, the OD period grows.

The degree of the period change depends on the decrease in the number of interocular connection,
and is difficult to estimate. A curious feature displayed by OD period in our model
is an abrupt increase at $N_s/N_o\approx 1.15$ by a factor of about $2.3$, cf. Fig.~\ref{fs}. 
This implies that close to point $N_s/N_o=1.15$ the OD period may be very sensitive to developmental manipulations.
This finding may have correlate in squirrel monkey, for which the observed increase in period is
by a factor $2.9-3.6$ (Horton and Hocking, 1996b). These data are obtained from comparison to a single strabismic animal.
The following scenario is possible, comparing squirrel monkey to the strabismus experiments
in owl monkey (Livingstone, 1996), in which no significant increase in periodicity is observed.
Parameter $N_s/N_o$ in squirrel monkey passes the point $N_s/N_o=1.15$ due to strabismus, leading to the substantial
increase in period. In owl monkey parameter $N_s/N_o$ is above $1.15$ in normal animal. Strabismus therefore
has little effect. This scenario is consistent with sharper OD columns in normal owl monkeys ($N_s/N_o>1.15$)
than in normal squirrel monkeys ($N_s/N_o<1.15$) (Livingstone, 1996; Horton and Hocking, 1996b). 
Experimentally induced strabismus in cat leads to an increase in the OD period by a factor of $1.3$ 
(Lowel, 1994; Goodhill, 1993; see however Jones et al., 1996). Our model suggests that parameter $N_s/N_o>1.15$ in cat,
and the increase in the period is due to the smooth part of the dependence in Fig.~\ref{fs},
which may not be so substantial as in squirrel monkey.

\subsection{On the importance of wiring minimization}

The relevance of wiring economy principle to the neuronal spatial organization
can be illustrated by the following thought experiment (Koulakov and Chklovskii, 2001). 
Imagine taking a cortical area and scrambling 
neurons in that area, while keeping all the connections between neurons the 
same. Since the circuit is unchanged, the functional properties of the neurons 
remain intact. Therefore, from the functional point of view, the scrambled 
region is identical to the original one. In fact, the only difference caused by 
scrambling is in the length of neuronal connections. Therefore, it is hard (if 
not impossible) to justify the existence of systematic cortical maps, such as OD pattern, 
without invoking the cost of making long neuronal connections. Although some theories of 
map formation may not explicitly mention the wiring optimization principle, it 
is present implicitly, usually in requiring the locality of intra-cortical 
connections.

How important is the constraint imposed by wiring minimization? In principle 
one can imagine development of an organism, which has 30\% excess of wire with 
respect to the existing ones. It turns out that the existence of such an organism is 
close to impossible. Indeed, imagine that an external object, such as a blood 
vessel, is introduced in certain area of the gray matter. In this case some of the 
neuronal connection have to go around the vessel, therefore increasing in 
length. If the nerve pulses are to be delivered at the original speed and/or 
intensity, the elongated axons and dendrites have to be made thicker, to 
increase the pulse propagation speed and decrease dendritic attenuation. This 
leads to more obstacles on the way of other neuronal connections and so on. 
Thus, introduction of a new blood vessel leads to an infinite series of axonal 
and dendritic reconstructions. The same is true about the extra connection 
volume, resulting from wasteful neuronal positioning. Such infinite series of 
reconstructions can diverge, which implies that the connection volume resulting 
from more and more reconstructions increases indefinitely. In this case the 
new blood vessel can never be inserted without sacrificing significantly the 
brain function. It turns out that mammalian brain has reached the verge of 
this so called {\it wiring catastrophe} (Chklovskii and Stevens, 2001), so that 
it gets increasingly more difficult to accommodate excess volume in the nerve 
tissues. 

The wiring catastrophe occurs when the fraction of axons and dendrites in the
cortical volume reaches $60\%$. Electron microscopy studies of 
cortical slices show that the actual volume occupied by neuronal processes 
is about $55\%$ (Chklovskii and Stevens, 2001).
The brain therefore has approached the barrier imposed by wiring catastrophe. 
Further increase of the volume fraction of neuronal processed may deteriorate
the brain function.

\subsection{Comparison to other models}

As discussed in the previous subsection, wiring optimization is the only known 
way to relate neuronal layout (as manifested in the ODP) to the statistics of 
neuronal connectivity. Models of the ODP development that do not explicitly rely 
on wiring optimization invoke it implicitly, usually requiring the locality of 
intra-cortical connections.

Here we discuss the relationship of our model to the models that invoke wiring 
optimization explicitly.

In his pioneering work, Mitchison (1991) studied a question whether ODP minimize the 
wiring volume relative to the {\it Salt and Pepper} layout. He assumed that the inter-
neuronal connectivity is determined both by ocular dominance and retinotopy with 
all neurons having the same connectivity rules. He found that the answer to this 
question depends on the detailed assumptions about axonal branching rules. 
In particular it depends on the value of axonal branching exponent $\alpha$.
He has shown that if all axonal segments have the same caliber ($\alpha=\infty$), 
than ODP's are indeed advantageous for certain range of ratios of same-eye to opposite-eye 
connections.  He also showed that if $\alpha<4$ than the ODP do not save wiring volume relative to {\it Salt and Pepper}.
However, existing data seems to suggest that axonal caliber branches with $\alpha<4$ 
(Deschenes and Landry 1980, Adal and Barker, 1965).

The case of axonal branching with the cross-sectional area conservation 
corresponds closely to our model because we require a separate connection for 
each neuron. The reason we find that ODP minimize wiring length is because we 
drop the retinotopy requirement on inter-neuronal connection rules, an 
assumption supported by the experimental data (Katz). Although, effectively 
connections are roughly retinotopic, connection rules may vary from neuron to 
neuron thus providing some flexibility. The advantage of our approach is its 
simplicity allowing us to map out a complete phase diagram and make 
experimentally testable predictions. The full theory of the ODP will require a 
detailed analysis of axonal branching which must rely on better knowledge of 
axonal branching rules.

Jones et al. (1991) proposed an explanation for why ODP have either Stripy or Patchy 
appearance. They assumed that neurons are already segregated into the ODP (by 
considering units whose size equals the width of monocular regions) and found 
that the difference between Stripy and Patchy appearances of the ODP could be 
due to the boundary conditions, i.e. different shape of V1 in different species. 
Although the correlation between the shape of V1 and the ODP layout is observed, 
the model of Jones et al. does not explain why peripheral representation of 
macaque V1 has patchy ODP or why ocular dominance stripes run perpendicular to 
the long axis of V1 in some parts of V1 but not in others. Moreover, it is the 
local structure of ODP that is likely to determine the shape of V1 and not the 
other way around. Therefore, unlike Jones et al., our work proposes a unified 
theory of ODP including {\it Salt and Pepper}, Stripy and Patchy layouts, and is based on 
local inter-neuronal connectivity rules.

Another model related to wiring length minimization is the elastic net model 
studied by Goodhill and coworkers (1993). The original formulation of the model 
minimized the cost function which penalized for placing nearby neurons whose 
activity is not correlated, a choice justified by computational convenience. 
Later the elastic model was generalized by the introduction of a C measure. 
Maximization of C measure effectively corresponds to penalizing for placing 
correlated neurons far apart. Unlike wiring optimization the penalty does not 
increase beyond a distance called cortical interaction. Because of this, elastic 
net often yields solutions where left and right eye neurons are completely 
segregated into left and right eye maps.

Our wiring optimization models can be viewed as a sub-set of models described by 
C measure. The advantage of our  wiring optimization approach is that it has a 
transparent biological justification for the cost of placing neurons far from 
each other - the cost of wiring. Because of this, wiring optimization is a 
natural choice for questions related to the anatomy of intra-cortical 
connections. 

Wiring optimization provides a link between neuronal connectivity and spatial 
layout. Thus, it leaves open the connection between connectivity and 
computational function. Unlike most other models, which assume that neurons should 
be connected only if they are correlated, wiring optimization makes other assumptions about 
connectivity possible, for example connecting neurons with anti-correlated firing.

\subsection{Conclusion}

Our theory relates functional requirements on the neuronal circuits to its structural properties. 
In particular, our model relates the amounts of the neuronal intraocular and interocular interactions, and
the filling fraction of ipsilateral neurons, to the structure of the ocular dominance pattern. 
We conclude that the interspecies variability in the ocular dominance patterns
may be explained by differences in the underlying cortical circuitry.



\end{document}